\documentclass{jpp}
\usepackage{physics}
\usepackage[utf8]{inputenc}
\usepackage[T1]{fontenc}
\usepackage{amsmath}
\usepackage{hyperref}
\usepackage[dvipsnames]{xcolor}
\usepackage{siunitx}

\graphicspath{figs/UpdatedOskarDraft/}

\shorttitle{Scrape-off losses of runaway electrons during tokamak disruptions}
\shortauthor{O. Vallhagen and others}

\title{Reduced modeling of scrape-off losses of runaway electrons during tokamak disruptions}
\author{O.~Vallhagen\aff{1}, L.~Hanebring\aff{1}, T.~F\"{u}l\"{o}p\aff{1}, M. Hoppe\aff{2}
 \and    I.~Pusztai\aff{1}}
\affiliation{\aff{1}Department of Physics, Chalmers University of Technology, G\"{o}teborg, SE-41296, Sweden
\aff{2}Department of Electrical Engineering, KTH Royal Institute of Technology, SE-11428 Stockholm, Sweden
}
\begin{document}

\maketitle

\begin{abstract}
Accurate modeling of runaway electron generation and losses during tokamak disruptions is crucial for the development of reactor-scale tokamak devices. In this paper we present a reduced model for runaway electron losses due to flux surface scrape-off caused by the vertical motion of the plasma. The model is made compatible with computationally inexpensive one-dimensional models averaging over a fixed flux-surface geometry, by formulating it as a loss term outside an estimated time-varying minor radius of the last closed flux surface. 
We then implement this model in the disruption modeling tool DREAM, and demonstrate its impact on selected scenarios relevant for ITER. 
Our results indicate that scrape-off losses may be crucial for making complete runaway avoidance possible even in a $15\,\rm MA$ DT H-mode ITER scenario. The results are however sensitive to the details of the runaway electron generation and phenomena affecting the current density profile, such as the current profile relaxation at the beginning of the disruption.
\end{abstract}

\section{Introduction}
One of the main open questions for the design of a tokamak power plant is how to mitigate the potential damage to the machine caused by \emph{disruptions} -- undesired off-normal events when the plasma confinement is suddenly lost \citep{Hender, Hollmann2015}. These may harm the machine in three main ways: through localized heat loads due to the rapid loss of the thermal energy during the first stage of the disruption, the thermal quench (TQ); magnetic forces associated with eddy currents and halo currents flowing (partly) through the conducting structures in the machine, generated during the current quench (CQ); and the formation of a beam of highly energetic runaway electrons (REs), which may cause damage down to the sub-surface of plasma facing components upon impact.

One of the main disruption mitigation techniques currently pursued is shattered pellet injection (SPI), which has been chosen as the baseline strategy for ITER \citep{LehnenITER}. This technique was originally developed and successfully tested in 2010 at the DIII-D tokamak \citep{Commaux2010, Commaux2016, Baylor2019, Combs2018}. Similar systems have since been installed at several other devices, including an upgraded system at DIII-D \citep{Meitner2017}, ASDEX Upgrade \citep{Dibon2023}, KSTAR\citep{Park2020} and JET \citep{Baylor2021, WilsonSOFT}, where they have provided experimental support for the design and operation of the ITER disruption mitigation system \citep{Herfindal2019, Papp2021, Jachmich2022, Baylor2021}.

There are however major differences in some of the machine and plasma parameters in ITER compared to present day devices, notably the machine size and plasma current. The scaling of the disruption mitigation performance with respect to these parameters is non-trivial, and some aspects of a reactor-relevant disruption mitigation system can therefore not be studied at present day devices. This creates a need for a reliable modelling capability. Advanced simulation tools exist which can model the full three-dimensional (3D) magnetohydrodynamic (MHD) evolution during a disruption, such as JOREK \citep{Huysmans2007MHD,Hoelzl_2021}. However, they are not suited for large parameter scans covering all phases of the disruption due to their extremely high computational expenses. Therefore, such explorations have instead been performed using reduced one-dimensional (1D) models averaging over a fixed magnetic flux surface geometry, and employing a simplified model for the TQ to emulate the effect of flux surface break-up. These include  starting directly from a post-TQ plasma \citep{Martin2017Formation}, mimicking the TQ with a prescribed temperature drop \citep{Vallhagen2020Runaway, Pusztai2023Bayesian, Ekmark2024Bayesian}, or prescribing strongly enhanced transport coefficients \citep{Vallhagen2024ITER} informed by 3D modeling results. Recent studies of the latter type indicate that the RE current cannot be kept below a few MAs in a $15\,\rm MA$ DT H-mode ITER scenario, while also satisfying the other disruption mitigation requirements \citep{Vallhagen2024ITER, Ekmark2024Bayesian}.

The simplifications made in the above mentioned studies might however result in unrealistically large RE currents. In particular, neglecting the vertical plasma motion, and subsequent plasma scrape-off, might lead to a substantial overestimation of the RE current; open field line regions cannot sustain a RE current, as the REs are lost to the wall. It has been shown by \cite{Wang2024Effect}, with two-dimensional (2D) axisymmetric simulations with the JOREK code, that the vertical plasma motion can substantially reduce the effective avalanche gain in ITER disruptions. The vertical plasma motion was found to significantly reduce the variation of the poloidal flux (between a given flux surface and the center of the torus) while the flux surfaces are intact. The maximum RE avalanche gain (before the RE current saturates) is exponentially sensitive to this flux variation \citep{BoozerHyperres}, potentially leading to an important impact on the final RE current. The importance of the poloidal flux available for RE generation was demonstrated numerically, for example by \cite{Vallhagen2024ITER} by artificially varying the boundary condition for the electric field. 

An accurate prediction of the RE generation in ITER should therefore account for RE losses associated with the vertical plasma motion. Moreover, it is desirable to have such capability in low computational cost 1D simulation tools, such as DREAM \citep{DREAM}. This paper therefore aims to develop a simplified model for the RE losses due to flux surface scrape-off in a 1D framework averaging over an otherwise fixed flux surface geometry. This is achieved by finding an approximate criterion determining which flux surfaces are scraped off, and prevent them from carrying a current, enforced by a loss term. This model is implemented in the DREAM code, and is demonstrated by simulating a selection of SPI-mitigated disruption scenarios of interest for ITER.

The rest of this paper is structured as follows. The reduced model for RE scrape-off losses is described in section \ref{sec:REModel}, and the disruption model and SPI scenarios considered are described in section \ref{sec:SPISetup}. Simulation results of the selected cases are presented in section \ref{sec:results}, comparing results with and without scrape-off losses included. The results are then discussed in section \ref{sec:discussion}, and the conclusions are summarized in section \ref{sec:conclusion}.


\section{Disruption model and scenarios}


\subsection{Runaway electron model with scrape-off losses}
\label{sec:REModel}
In this section we describe the reduced model of RE losses due to flux surface scrape-off employed in this work. This model builds on the observation in JOREK simulations \citep{Wang2024Effect} that the change in the poloidal magnetic flux during the CQ at the instantaneous last closed flux surface (LCFS)is small. This can be seen in figure 4 in \cite{Wang2024Effect}, showing that the poloidal flux at the LCFS only varies by a few $\rm Wb$\footnote{The poloidal flux at the LCFS might increase notably during the RE plateau phase, as shown in figure 5 in \cite{Wang2024Effect}. This increase is however still rather small compared to the initial flux at the core, and the increase only occurs when a substantial RE beam has already formed. The flux at the LCFS can thus be considered essentially constant during the majority of the RE generation phase.}, which is much smaller than the initial poloidal flux of $\sim 80\,\rm Wb$ at the plasma core. 

A qualitative explanation for this observation can be found by considering that the tokamak wall is a good conductor on the CQ time scale (the resistive time scale of the first toroidally closed structure in ITER is about $500\,\rm ms$ \citep{Vallhagen2024ITER}). The high conductivity of the wall prevents magnetic flux from diffusing through the wall and keeps the flux in the vicinity of the wall nearly constant on the CQ time scale. As the plasma stays relatively close to the conducting wall (compared to the characteristic length scale of variation of the poloidal flux along the closest distance between the plasma and the wall, as exemplified by figure 1 in \cite{Wang2024Effect}), on such a time scale, the flux at the LCFS remains essentially constant.

This observation allows us to distinguish between closed and open field line regions in a model with an otherwise static flux surface geometry. It should be clarified here that although the plasma geometry at a given flux surface label $r$, defined as the distance between the magnetic axis and
the flux surface along the outboard mid-plane, is held fixed, one may still  calculate a separate self-consistent poloidal flux, related to the instantaneous plasma current density via Amp\`{e}re's law. This self-consistent poloidal flux, denoted $\psi _\mathrm{p}$, may be used to estimate the radius of the actual LCFS; flux surfaces with $\psi _\mathrm{p}$ larger\footnote{Note that we define the poloidal magnetic flux such that it increases from the conducting wall inward towards the plasma center.} than its initial value at the plasma edge are considered closed, while the field line region outside is considered open.
The open field line region is not allowed to carry any REs, as they are lost to the wall very rapidly (on a time scale $\sim R/c\sim 10^{-8}\,\rm s$, where $R$ is the tokamak major radius and $c$ is the speed of light) and lose their kinetic energy there. A simple and numerically tractable way to enforce this is to introduce a loss term to the RE density $n_\mathrm{RE}$ on a (static) flux surface labeled $r$ in the model according to
\begin{equation}
    \left(\frac{\partial n_\text{RE}}{\partial t}\right)^\text{scrapeoff} = -\frac{n_\mathrm{RE}}{t_\mathrm{loss}}\Theta (r-r_\mathrm{LCFS}),
\end{equation}
where the flux surface label $r_\mathrm{LCFS}$ for what is considered the LCFS is determined by (numerically) solving
\begin{equation}
    \psi _\mathrm{p}(r_\mathrm{LCFS}) = \psi _\mathrm{p}(a, t=0).
\end{equation}
Here, $\Theta$ is the Heaviside step function, $a$ is the (initial) plasma minor radius, $b$ is the radius of the conducting wall and $t_\mathrm{loss}$ is the time scale for the loss in the open field line region. 

A physically motivated choice for the loss time scale is to set $t_\mathrm{loss}\sim R/c$. This time scale is however typically several orders of magnitude faster than any other time scale of interest (except for short phases during the disruption when a rapid ionization or recombination process may occur). One may therefore set $t_\mathrm{loss}\sim 10 \Delta t$, as some practical choice, where $\Delta t$ is the numerical time resolution, in order to avoid increasing the need for time resolution without altering the results.


\subsection{Scenarios and simulation setup}
\label{sec:SPISetup}
With the above model at hand, we now turn to the details of the SPI scenarios used to study the effect of RE scrape-off losses with the DREAM code. The RE dynamics in ITER is expected to differ significantly with and without the presence of nuclear RE generation mechanisms, such as $\beta^{-}$-decay and Compton scattering of $\gamma$-photons from the radioactive wall \citep{Vallhagen2024ITER}. We therefore consider one case with a pure hydrogen (H) background plasma, referred to as scenario 1, and one case with an equal mix of deuterium (D) and tritium (T), referred to as scenario 2. The chosen scenarios both have an initial plasma current of $15\,\rm MA$, giving the most challenging condition for RE mitigation in ITER, while they differ in their SPI settings as well as their initial temperature and density; scenario 1 is an L-mode case with a core temperature of $\sim 5\,\rm keV$ and density $\sim 5\cdot 10^{19}\,\rm m^{-3}$, while scenario 2 is an H-mode case with a core temperature of $\sim 20\,\rm keV$ and density $\sim 8\cdot 10^{19}\,\rm m^{-3}$. 

Both scenarios are produced with the CORSICA code \citep{CORSICA}, and have previously been studied from a disruption mitigation perspective in \cite{Vallhagen2024ITER}, where they are referred to as M4 and St4 in appendix B, respectively. Scenario 1 was chosen as a representative case from the pre-nuclear phase of ITER operation with a mediocre predicted RE mitigation performance, and scenario 2 was chosen as it was the best performing nuclear case. Further details of the scenarios may also be found in \cite{Vallhagen2024ITER}, while  the specifics relevant for the present study are given below.


\subsubsection{SPI model}
Both scenarios represent a disruption mitigated by a mixed D-Ne SPI. In scenario 1, three pellets of the size planned for ITER are injected simultaneously, containing $1.85\cdot 10^{24}$ atoms out of which $2\cdot 10^{23}$ are Ne. Scenario 2 starts with the injection of a pure D pellet, followed after $5\,\rm ms$ by a mixed pellet with $2.5\cdot 10^{22}$ Ne atoms. All pellets considered in this paper are assumed to be shattered into 487 shards with a size distribution given by the fragmentation model by \cite{ParksShattering}, as in previous simulations with the INDEX code \citep{AkinobuITPA}. The shards are initiated on a point $(R, Z) = (8.568, 0.6855)$, where $R$ denotes the major radius coordinate and $Z$ denotes the height above the midplane. They are then assumed to travel through the plasma with an average speed of $v_\mathrm{p}=500\,\rm m/s$, a uniform speed distribution between $v_\mathrm{p}\pm \Delta v_\mathrm{p}$ with $\Delta v_\mathrm{p}/v_\mathrm{p}=0.4$, and a divergence angle of $\alpha = \pm 10^\circ$.

Once the shards enter the plasma, they ablate according to the Neutral Gas Shielding (NGS) model \citep{ParksTSDW}. The material is deposited in its neutral state, and the density of each of the ionization states is then evolved using time dependent rate equations. The ionization and recombination rates are taken from the ADAS database \citep{ADAS} for Ne and from the AMJUEL database\footnote{http://www.eirene.de/html/amjuel.html} for D. The latter includes the effect of opacity to Lyman radiation, which has been shown to play an important role in the ionization and energy balance following a disruption mitigated by SPI with a large D content \citep{VallhagenTwoStage}.

For the Ne-doped shards, the material is deposited directly where it is ablated. However, for pure D pellets, the much weaker radiation from the cool plasmoids building up around the pellet shards allows the pressure in the plasmoids to become much higher than for the Ne-doped shards. This makes the plasmoids much more prone to drift outwards along the major radius, due to the $E\times B$-drift resulting from the charge separation caused by the $\nabla B$-current inside the plasmoid \citep{ParksDrift, PegourieDrift, VallhagenDrift}. Such a drift may significantly shift the deposition profile, possibly resulting in material being ejected from the plasma. Moreover, it also shifts the cooling of the background plasma away from the shards, making them ablate faster and deposit their material earlier along their trajectories. To account for this effect, we deposit the material ablated from the pure D pellets 2 radial grid cells behind the shards, corresponding to $\sim 20\,\rm cm$ with the current resolution. 

We note that the drift displacement may be significantly longer than $20\,\rm cm$ under some relevant circumstances \citep{VallhagenDrift}, so that this assumption may give a rather optimistic result for the material assimilation rate. The assumed displacement is however sufficient to result in a rather edge-localized deposition profile here, partly due to the aforementioned displacement of the background plasma cooling.


\subsubsection{TQ model}
\label{sec:TQModel}
As the SPI shards pass through the plasma it will rapidly cool, partly due to radiation from the injected material and partly due to the stochastisation of the magnetic field as the plasma becomes MHD unstable. The radiation losses are modeled using rate coefficients from the ADAS and AMJUEL databases, corresponding to the ionization and recombination rates mentioned above. As DREAM does not self-consistently model the MHD activity in the plasma, the stochastisation of the magnetic field is accounted for by employing enhanced transport coefficients, triggered based on two different criteria. For the heat, we impose a diffusive transport of Rechester-Rosenbluth type \citep{RechesterRosenbluth}. The magnetic perturbation level is set so that the temperature would drop to around $200\,\rm eV$ during the transport event with transport as the only loss mechanism\footnote{The perturbation level necessary to yield the mentioned cooling rate is determined from separate simulations.}. At this point, the transport along the stochastic field lines becomes relatively inefficient, and the losses are instead dominated by radiation. For the ions we use a combination of diffusion and advection coefficients to emulate the rapid MHD mixing observed in 3D simulations with the JOREK code \citep{Hu2021}. 

In scenario 1, the transport event is triggered when a shard with speed $v_\mathrm{p}$ passes the flux surface characterized by the safety factor $q=2$. At this point, the local plasma cooling around the shards may be sufficient to perturb the plasma pressure and current density enough to trigger a rapidly growing MHD instability. The duration of the transport event is assumed to be $t_\mathrm{TQ}=1\,\rm ms$, which is achieved by using a normalized magnetic perturbation amplitude of $\delta B/B = 3.74\cdot 10^{-3}$. This results in a relatively early and short TQ, providing relatively challenging conditions for RE avoidance. In scenario 2 on the other hand, which was designed to be more optimistic (although still plausible), the transport event is triggered when the flux surface averaged temperature falls below $10\,\rm eV$ anywhere inside the $q=2$ surface. The duration of the transport event is here assumed to be $t_\mathrm{TQ}=3\,\rm ms$, corresponding to $\delta B/B=2.01\cdot 10^{-3}$. 

In all cases presented in this work, the ion diffusion coefficient is set to $D_\mathrm{ion}=4000\,\rm m^2/s$ at the start of the transport event, and the ion advection coefficient is set to $A_\mathrm{ion}=-2000\,\rm m/s$. From this point in time both coefficients decay exponentially with an e-folding time of $t_\mathrm{ion}=0.5\,\rm ms$, and are thus active over a time scale comparable to the heat transport. A similar form of the ion transport coefficients has previously been used to reproduce AUG disruption experiments with the ASTRA code \citep{Linder}. The values used here are adapted to result in a mixing on a similar time scale as found in 3D JOREK simulations of ITER by \cite{Hu2021}. 


\subsubsection{Current evolution}

When the plasma reaches temperatures around $10\,\rm eV$, the plasma current starts to decay, and in the meantime a strong electric field is induced, which may generate REs. The current density profile is typically also significantly flattened during the transport event, leading to a temporal spike in the total plasma current. To account for this, we solve the mean field equation for the self-consistent poloidal flux $\psi _\mathrm{p}$ including a hyperresistive term \citep{BoozerHyperres, PusztaiHyperres} during the transport event. The strength of the hyperresistive term is determined by the hyperresistivity parameter $\Lambda _\mathrm{m}$, which we set to $\Lambda _\mathrm{m}=0.1\,\rm Wb^2m/s$ in order to get a significant flattening of the current density profile and a reasonable size of the spike in the total plasma current.

The current density is divided into an ohmic part and a fluid-like RE part consisting of REs assumed to travel at the speed of light. The REs are modeled using the fluid model available in DREAM \citep{DREAM}. In scenario 1, only non-activated RE sources are present: the Dreicer mechanism, which is modeled using a neural network trained on the output from kinetic simulations \citep{HesslowNN}; the acceleration of the weakly collisional hot tail of the initial distribution function, modeled as described in appendix C in \cite{DREAM}; and the exponentiation of an existing seed of REs by the avalanche mechanism, accounting for effects of partial screening and radiation losses, as derived in \cite{HesslowAvalanche}. In scenario 2, an additional RE seed is generated by the tritium $\beta^{-}$-decay and Compton scattering of $\gamma$-photons from the activated wall \citep{FulopElongation, Martin2017Formation}. As noted in \cite{Martin2017Formation}, the $\gamma$-photon flux is expected to drop by a factor $\sim 1/1000$ when the fusion reactions, and hence the neutron bombardment of the wall, stops. We therefore use the nominal $\gamma$-photon flux in ITER as given in \cite{Martin2017Formation} until the end of the transport event, and then decrease the $\gamma$-photon flux by a factor $1/1000$.

Once generated, REs can be lost in three ways: slowing down as the electric field goes below the critical field for RE generation, accounted for by letting the avalanche growth rate become negative at such electric fields \citep{HesslowAvalanche}; scrape-off losses, modeled as described in section \ref{sec:REModel}; and diffusive transport during the transport event. For the latter, we have assumed a momentum dependent transport coefficient of the form $D_0 p/(1+p^2)$, where $p$ is the electron momentum normalized to $m_\mathrm{e} c$, with the electron mass $m_\mathrm{e}$. $D_0$ is taken to be the Rechester-Rosenbluth diffusion coefficient for an electron traveling at the speed of light, with the same magnetic perturbation amplitude as given in section \ref{sec:TQModel} for the heat transport. This momentum dependence captures the linear behavior expected at low speeds, while also accounting for the reduction due to averaging of the small-scale magnetic perturbations over finite orbit widths experienced at high energies. The momentum dependent diffusion coefficient is turned into the required diffusion coefficient for the total fluid RE density, by integrating the evolution equation for the RE distribution over momentum space, assuming a separation of the RE generation and transport scale, as detailed in \cite{SvenssonTransport}.


\section{Simulation results}
\label{sec:results}

We now turn to the simulations of the two disruption scenarios described in section \ref{sec:SPISetup}, focusing on the RE current density evolution with and without scrape-off losses. 

\begin{figure}
\centering
\includegraphics[width=\textwidth]{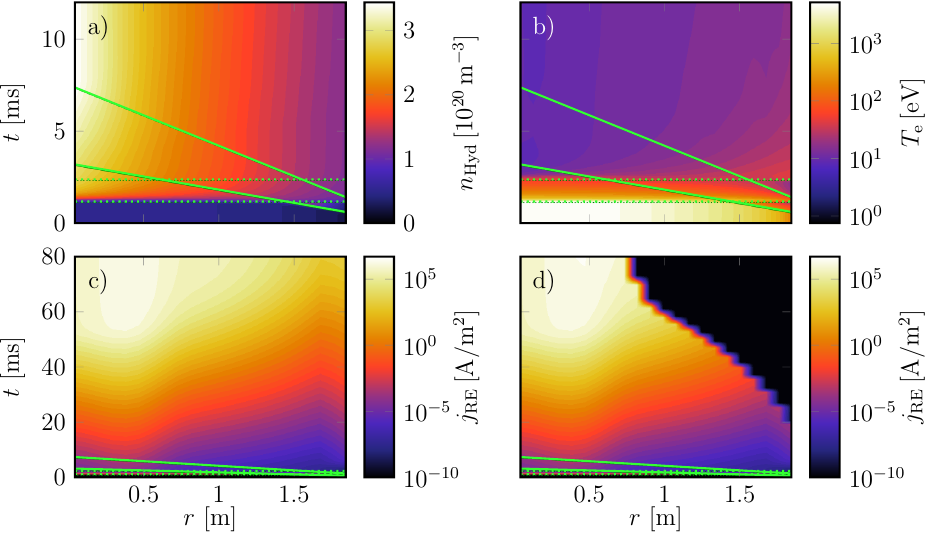}
\includegraphics[width=0.49\textwidth]{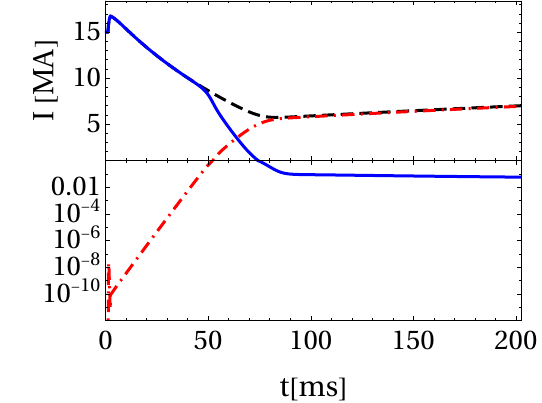}
\put(-20,125){\small e)}
\includegraphics[width=0.49\textwidth]{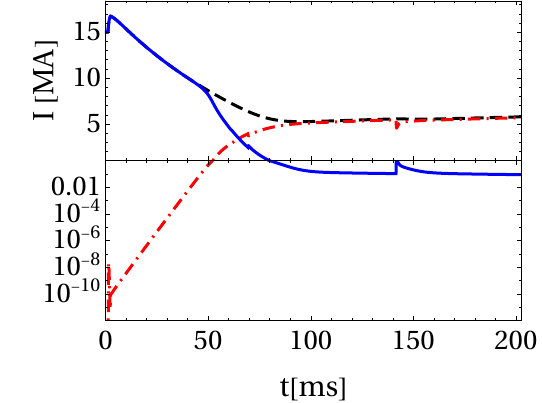}
\put(-20,125){\small f)}
\caption{Evolution of plasma profiles for scenario 1; hydrogen density (a), temperature (b) and RE current density evolution without (c) and with (d) scrape-off losses. The solid green lines indicate the trajectories of the fastest and slowest pellet shards, and the dotted green lines indicate the time span of the transport event. Integrated ohmic (solid), RE (dash-dotted) and total (dashed) currents are also shown without (e) and with (f) scrape-off losses (note that up to $1 \,\rm MA$ the y-axis is logarithmic).}
\label{fig:M4}
\end{figure}

Figure \ref{fig:M4} shows the evolution of the hydrogen density, electron temperature, and RE current density with and without scrape-off losses for scenario 1 (the hydrogen density and electron temperature profiles are similar with and without scrape-off losses). Due to the relatively early and fast TQ, the pellet material has rather little time to ablate before the plasma temperature falls below $\sim 100\,\rm eV$, at which point the ablation rate is drastically decreased, as shown in figures \ref{fig:M4}a-b). As a result, the material assimilation rate is only $1.89\,\rm \%$. Although the post-TQ temperature is rather low and the ablation slow, due to the low overall assimilation rate the ablation taking place after the transport event is sufficient to continue to notably alter the density profile. At this point the pellet shards are close to the plasma core, where the flux surfaces are smaller. This results in a density profile that is somewhat peaked in the core, and consequently a hollow temperature profile, which is not flattened by the transport event. 

Most of the RE seed electrons generated during the transport event are lost via diffusion, but there is a remaining effective RE seed current of $0.14\,\rm m A$, defined as the sum of the RE current at the end of the transport event and the seed growth rates integrated over the rest of the CQ. Moreover, the low assimilation rate allows the avalanche mechanism to amplify the RE current for a relatively long time before the electric field drops below critical. Without scrape-off losses, this is sufficient to generate a RE plateau current of $5.43\,\rm MA$ (see figure \ref{fig:M4}e). This representative value of the  RE current is defined as the RE current at the moment the RE fraction of the total current just reaches 95\%. The current density is localized to the inner parts of the plasma where the temperature is lower (note the logarithmic color scale), as shown in figure \ref{fig:M4}c).

With scrape-off losses included, some REs in the outer parts of the plasma are initially lost as the plasma current, and hence the poloidal flux at a given radius, decays, giving rise to the black area in figure \ref{fig:M4}d), void of REs. However, since most of the RE current is generated in the inner part of the plasma, these initial RE losses are rather small, allowing the RE current to continue growing largely unaffected. By the time the LCFS approaches the inner parts of the plasma, the RE current is sufficiently large to significantly slow the decay of the total plasma current, thus also slowing the decay of the poloidal flux and the inwards motion of the LCFS. Eventually, a RE plateau is reached, which essentially halts the decay of the poloidal flux -- along with the inwards motion of the LCFS -- before the central flux surfaces can be scraped off, allowing the RE plateau to remain stable. Thus, in this example, the representative RE plateau current is only moderately reduced compared to the case without scrape-off losses, to $4.99\,\rm MA$ as shown in figure \ref{fig:M4}f).

\begin{figure}
\centering
\includegraphics[width=\textwidth]{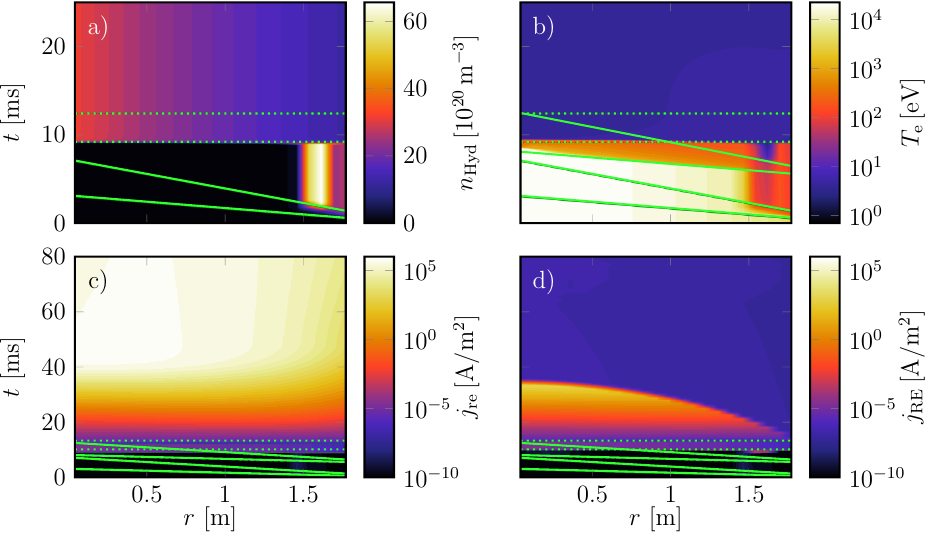}
\includegraphics[width=0.49\textwidth]{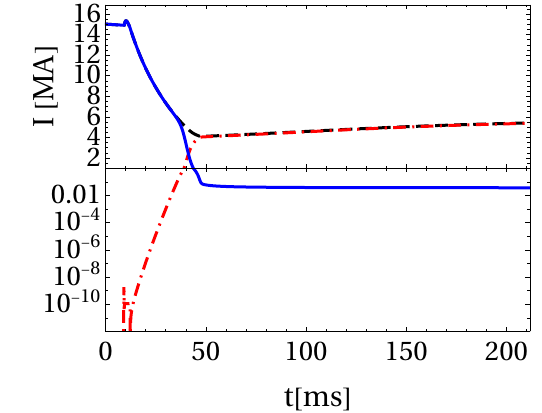}
\put(-20,125){\small e)}
\includegraphics[width=0.49\textwidth]{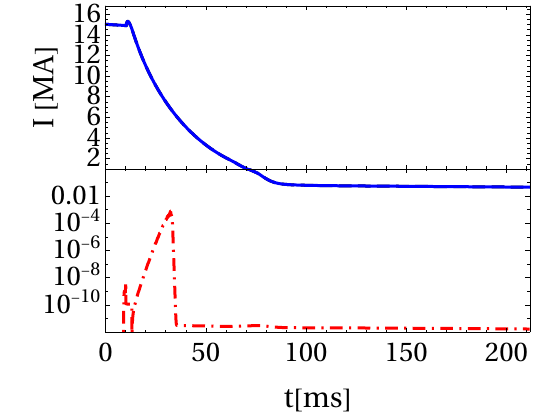}
\put(-20,125){\small f)}
\caption{Evolution of plasma profiles for scenario 2; hydrogen density (a), temperature (b) and RE current density evolution without (c) and with (d) scrape-off losses included. The solid green lines indicate the trajectories of the fastest and slowest pellet shards for each pellet, and the dotted green lines indicate the time span of the transport event.  Integrated ohmic (solid), RE (dash-dotted) and total (dashed) currents are also shown without (e) and with (f) scrape-off losses (note that that up to $1 \,\rm MA$ the y-axis is logarithmic).}
\label{fig:St4}
\end{figure}

The simulation results for scenario 2 differ in several aspects from scenario 1. In scenario 2, the first injection stage, consisting of pure D, passes completely without triggering the transport event, as shown in figure \ref{fig:St4}a). The radiation losses following a pure D injection also remain very small, so that the plasma is only cooled by dilution (see figure \ref{fig:St4}b) and the ablation rate stays relatively high during this injection stage. The assimilation rate of this injection stage is thus $63.5\,\rm\%$, i.e.~much higher than for scenario 1. Due to the drift-induced shift of the deposition for pure D pellets, the deposition profile is now more edge-localized, so that the temperature in the central parts of the plasma remains high even after this injection stage. This allows for a high assimilation rate of $57.6\,\rm\%$ also for the second injection stage. The second injection leads to a temperature drop to $\lesssim 5\,\rm eV$ throughout the entire plasma, with a rather flat profile.

The high assimilation rate, combined with the relatively late and slow transport event, effectively suppresses the non-activated RE seed mechanism. It also makes the CQ faster, due to the additional plasma cooling, and correspondingly increases the induced electric field. On the other hand it also increases the critical electric field, which helps to dampen the avalanche. The activated seed mechanisms however produce a seed of $2.3\,\rm mA$, which is notably larger than in scenario 1. The representative plateau RE current without scrape-off losses therefore still reaches a macroscopic value of $3.93\,\rm MA$, as shown in figure \ref{fig:St4}e).

Even though the RE current in scenario 2 (without scrape-off losses) is only moderately reduced compared to scenario 1, this difference has an important effect when scrape-off losses are activated. The poloidal flux is also affected by the RE current density profile being wider (see figure \ref{fig:St4}c), due to the flatter density and temperature profiles. These differences both contribute to a more rapid drop of the poloidal flux, making the LCFS move further inwards and at a faster rate, allowing the REs to be lost at an earlier stage of the scrape-off process. As a result, all flux surfaces are now scraped off before an RE plateau current is formed, as shown in figure \ref{fig:St4}d). In this case the flux surfaces are scraped off already before the RE current reaches macroscopic values; the maximum RE current is only $576\,\rm A$, as shown in figure \ref{fig:St4}f).

\begin{figure}
\centering
\includegraphics[width=0.57\textwidth]{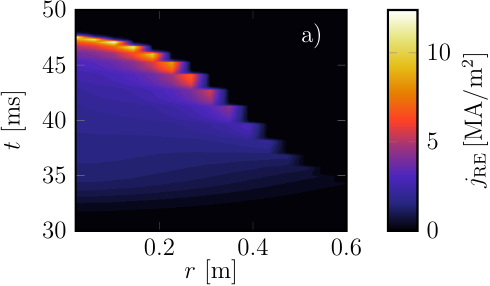}
\includegraphics[width=0.42\textwidth]{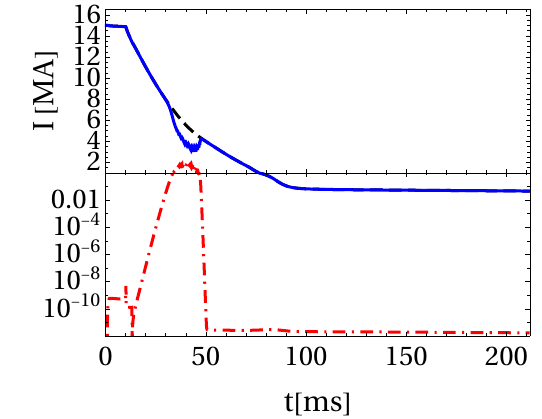}
\put(-20,110){\small  b)}
\caption{RE current evolution in scenario 2 without hyperresistivity. a) RE current density evolution, b) total RE current in the plasma (red dash-dotted) along with the total ohmic current (blue solid) and their sum (black dashed), c) total RE current lost due to scrape-off.}
\label{fig:St4NoHyperres}
\end{figure}

The scraped-off RE current may however be significantly affected by the details of the scenario and model parameters that can affect the decay of the poloidal flux, such as hyperresistivity. This is evident from figure \ref{fig:St4NoHyperres}, showing the RE current evolution in scenario 2 in the absence of hyperresistivity. All flux surfaces are still scraped off before a long-lived RE plateau is formed in this case, but only after the RE current has reached macroscopic values, peaking at $\sim 1.96\,\rm MA$ as shown in figure \ref{fig:St4NoHyperres}b). As the scraped-off RE current density is significant compared to the total current density, a strong electric field is induced just outside the LCFS (corresponding to the halo region in our model) as the total current density can not change faster than the resistive time scale. The electric field diffuses inwards to the closed flux surfaces, increasing the RE current density there. This mechanism gives rise to the skin current effect observed in figure \ref{fig:St4NoHyperres}a) (note that the color scale is now linear, to make this effect more visible), which is similar to the skin current observed in previous JOREK simulations \citep{Wang2024Effect}.


\section{Discussion}
\label{sec:discussion}
Comparing the behaviour of scenario 1 and 2 indicates that even a moderate difference in the RE current without scrape-off losses may lead to a major difference when scrape-off losses are included. This can be explained by a feedback effect between the RE generation and the scrape-off process: a slower RE generation allows the poloidal flux to drop further and faster, which leads to more flux surfaces being scraped-off at a higher rate, which reinforces the decrease of the RE current. Thus, while the scrape-off losses according to the present model may be small in some cases, they may be dramatic in other cases, with a rather sharp transition in between. 

In particular, a critical situation occurs when the RE generation becomes slow enough to allow for all flux surfaces to be scraped off before an RE plateau is formed. In ITER, this appears to happen at an RE current without scrape-off in the $4$-$6\,\rm MA$-range, but may also depend on other conditions affecting the poloidal flux, such as the hyperresistivity and other parameters that may affect the current profile. An accurate calculation of the RE current impacting the wall is therefore dependent on accurate estimates of these parameters. Scenario 2 without hyperresistivity also indicates the plausibility of intermediate cases where all flux surfaces are scraped off, but the scraped off RE current is still significant. 

Nevertheless, our results indicate that scenarios may exist where all flux surfaces are scraped off before the RE current reaches macroscopic values in an ITER $15\,\rm MA$ DT H-mode plasma. Thus, scrape-off losses could make a full RE avoidance in such a plasma possible, which previous studies (without scrape-off losses) indicated may not be the case \citep{Vallhagen2024ITER}. The scenarios considered here may even be rather conservative, as they do not account for any transport of REs along stochastic field lines during the CQ. It has previously been shown that such transport may have an important direct effect on the RE current, if the magnetic perturbation amplitude is high enough and covers the whole plasma \citep{SvenssonTransport}. With scrape-off losses included, the effect may be even larger as the extra losses would contribute to the positive feedback effect mentioned above. Moreover, as the plasma shrinks, the radial distance the REs have to diffuse over to reach the plasma edge decreases, leading to a reduced loss time scale. The shrinking of the plasma may also make it easier for externally induced magnetic perturbations (e.g., those induced by resonant magnetic perturbation coils) to reach the plasma core.

Scenario 2 without hyperresistivity illustrates that scraping off a macroscopic RE current may lead to the formation of a skin current, driven by energy diffusing from the open field line region to the closed flux surfaces. This effect was observed also in the 2D axisymmetric JOREK simulations by \cite{Wang2024Effect}, and could potentially increase the amount of REs impacting the wall. The amount of energy able to diffuse back into the closed flux surfaces however depends on the dissipation of energy to the first wall, by the REs as well as the Ohmic halo current. Such dissipation is not accounted for in the present model, and the skin current found here may thus be exaggerated. The interaction with the first wall can also affect other quantities, such as the temperature and density in the open field line region. To accurately account for this plasma-wall interaction however requires a detailed model that is outside of the scope of this paper.

The scrape-off losses at the end of the CQ may also be affected by the assumption of a constant poloidal flux between the wall and the plasma edge, which starts to become invalid at this point. Over a time scale longer than the CQ, the plasma can not be considered a good conductor, and while the flux variation is still limited by the higher conductivity of the wall, it was observed by \cite{Wang2024Effect} that the poloidal flux at the LCFS increases notably during the RE plateau phase (if a RE plateau is formed). The present model is therefore suitable to study the current dynamics during the RE generation process, but may be less accurate during the final loss of the RE beam.

There may also be intrinsically higher-dimensional effects which may affect the RE current evolution but can not be captured by the present model. One such effect is the non-uniform distance between the plasma and the wall at different poloidal angles, which may affect the inductive coupling and energy exchange between the plasma and the wall and give rise to 2D features of the plasma profiles, especially in the open field line region. The present model does also not capture changes in the shapes of the flux surfaces over time. Moreover, it was found by \cite{Wang2024Effect} that the RE current may differ notably between an upwards or downwards vertical displacement, while the present model can not distinguish between a displacement moving upwards or downwards. Thus, the present model should not be regarded as a suitable substitute for higher-dimensional models, but rather as a computationally efficient complement that allows results to be obtained faster and large parameter spaces to be scanned. Such parameter scans could then  be used to guide the efforts made with higher dimensional models.


\section{Conclusion}
\label{sec:conclusion}
We have developed a reduced model to estimate the scrape-off losses of runaway electrons due to the vertical plasma motion during a tokamak disruption. The model is based on the observation in 2D JOREK simulations that the poloidal flux between the tokamak wall and the instantaneous last closed flux surface remains approximately constant in time. This may be used to distinguish closed from open field line regions in an otherwise 1D framework with a fixed flux surface geometry. Our model enables scrape-off runaway electron losses to be accounted for in computationally inexpensive models, simplifying the exploration of this phenomenon in a wide parameter space. It may also be used to identify cases of interest to study with more complex and computationally expensive tools. 

Our results indicate that scrape-off losses may be crucial for making complete RE avoidance in an ITER $15\,\rm MA$ DT H-mode plasma possible. The results are however strongly sensitive to what the runaway current would be without the scrape-off losses, and also show a significant sensitivity to other conditions, such as the current profile relaxation during the thermal quench. This motivates further sensitivity studies and more refined simulations of scenarios similar to the promising scenario found in this work.


\section*{Acknowledgements} 
The authors are grateful to  E Nardon, C Wang,  P Helander and S Newton for fruitful discussions. This work was supported by the Swedish Research Council (Dnr.~2022-0286 and 2021-03943).The work has been carried out within the framework of the EUROfusion Consortium, funded by the European Union via the Euratom Research and Training Programme (Grant Agreement No 101052200 — EUROfusion). Views and opinions expressed are however those of the author(s) only and do not necessarily reflect those of the European Union or the European Commission. Neither the European Union nor the European Commission can be held responsible for them.

\section*{Declaration of Interests}
The authors report no conflict of interest.

\bibliographystyle{jpp}
\bibliography{bibliography}

\end{document}